\documentclass[]{article}

\usepackage{amssymb,amsmath}
\usepackage{hyperref}
\usepackage{booktabs}
\usepackage{graphicx}
\usepackage{subfigure}
\usepackage{float}
\usepackage{mwe}

\hypersetup{breaklinks=true}
\usepackage{fullpage}

\begin{document}
\begin{center}
    \Large
    \textbf{Longitudinal analysis of heart rate variability as it pertains to anxiety and readiness}
        
    \vspace{0.4cm}
    \large
    An exploratory single-case study
        
    \vspace{0.4cm}
    \textbf{Tucker Paron}
       
    \vspace{0.9cm}
    \textbf{Abstract}
\end{center}

\noindent The aim of this study is to explore the relationship between lifestyle choices, subjective experiences and objective biometric data in a single individual. The participant, at the time a male in his twenties, used the EliteHRV app to perform Heart Rate Variability Readings across 26 months accompanied by logs about the previous days activity as well as current emotional and physical state. The study will use a mixed-methods approach to analyze the data, including quantitative analysis of the biometric data and correlation analysis between the biometric data and subjective experience tags. Qualitative analysis of the daily logs will also be conducted to gain a deeper understanding of the participants' experiences and to identify keywords, people, or ideas that affect biometric output. The results of this study will provide insights into the relationship between subjective and objective measures, and the potential benefits or drawbacks of certain lifestyle choices and ways of thinking. The findings could have implications for the development of wearable-based personalized interventions for improving mental health and well-being. 

\section{Introduction}\label{sec:intro}
\subsection{Data}\label{subsec:data}
Data recorded with the \href{https://elitehrv.com}{EliteHRV} app using the \href{https://www.powr-labs.com/collections/all}{Powr Labs} Heart Rate Strap will be utilized. EliteHRV, now known as Spren, is a private North Carolina company that specializes in personal health. Their main product is an app that enables users to pair a heart rate monitor to measure and analyze their heart rate variability. The data used in this study has been collected from the EliteHRV app (using the Powr Labs HR strap) using the 'morning readiness readings' feature. These readings are taken within 5 minutes of waking and were recorded daily from December 26, 2020 to May 3, 2023, with the exception of November and December of 2021. For further information, please refer to the \hyperref[sec:discussion]{Discussion} section.

\noindent The daily readings, of which there are 747, consist of 14 \hyperref[subsubsec:biometric]{Biometric} variables (including HRV), 3 self-recorded physical and mental measures, 60 self-defined and self-recorded binary "Tags", a self-recorded string variable consisting of a "daily note", and a Morning Readiness score (1-10) generated by the app. All self-recorded variables are described in the \hyperref[subsubsec:self]{Self-measured} section.

\subsection{Purpose}\label{subsec:purpose}
This study is exploratory in nature, so it is inherently open-ended with the only real focus being on finding new patterns or trends to inspire future research. That being said, this study does aim to explore and answer three main questions:
\begin{enumerate}
    \item Determine the most significant correlates with extreme HRV and Readiness scores
    \item Identify linguistic trends in relation to HRV and Readiness scores
    \item Examine connection between anxiety and HRV
\end{enumerate}

\noindent The final question, about anxiety and HRV is particularly important; however to address anxiety and its relationship to human autonomic responses, it is crucial to understand its evolutionary function: "to prepare the individual to detect and deal with threats". Anxiety is an biological function evolved to "maximize survival and reproduction" (Bateson, 2011). That being said, individuals can have anxiety that is disorder adapted (ie. no longer serving to promote survival, but rather lowering quality of life). This study does not seek to diagnose the participant or any other individual with normal or abnormal anxiety mechanisms, but rather aims to observe the mechanisms and their (potentially) associated biometrics.

\noindent While this is a single-case study, the intention of this research is to challenge or support existing research about heart rate variability and the autonomic nervous system as it pertains to anxiety so as to better inform individuals' various lifestyle choices.

\section{Methods}\label{sec:methods}
\subsection{Study Design}
This study is a mixed-methods exploratory single-case study that aims to analyze the relationship between lifestyle choices, subjective experiences, and objective biometric data in a single individual. The study utilizes both quantitative and qualitative methods to analyze the data.

\subsection{Participants}
The participant (n=1) is male (early twenties) who has been using the EliteHRV app to perform Heart Rate Variability readings since December 26, 2020. He has provided consent to use his data for research purposes. The participant has been keeping a daily log of his activities, emotions, and physical state, which will be used for the qualitative analysis. The logging of the individual’s data was not supervised. 60 of the variables, including the binary Tags and the daily Notes section were both created and continually logged based on the participant’s judgment. Not all Tags were created on or before the first reading; however, most Tags were initialized within the first week of recording data.

\subsection{Data Collection}
The data used in this study has been collected from the EliteHRV app using the Powr Labs HR strap. The data consists of 747 daily readings recorded daily from December 26, 2020, to May 3, 2023, with the exception of November and December of 2021. The readings are taken within 5 minutes of the participants waking, lying down in whatever sleeping arrangement the participant resided in. The participant elected to read during these readings, which, given this was consistent across all readings, is acceptable. The readings consist of 14 Biometric variables (including HRV), 3 self-recorded physical and mental measures, 60 self-defined and self-recorded binary "Tags", a self-recorded string variable consisting of a "daily note," and a Morning Readiness score (1-10) generated by the EliteHRV app. All self-recorded variables are described in the \hyperref[subsubsec:self]{Self-measured} section.

\subsection{Quantitative Analysis}
The quantitative analysis will include calculating the correlation coefficients between the biometric variables and the subjective experience tags. The correlation coefficients will be calculated using three different methods depending on the variables in question. When observing two continuous variables, the standard Pearson correlation coefficient will be determined. This is due to the assumptions of a linear relationship and normal distribution. Conversely, if looking at a binary and a continuous variable, the Point Biserial Correlation coefficient will be utilized. Lastly, if both variables are binary, the designated method will be Chi-squared contingency as this is specific to categorical variables. The quantitative analysis will also include the generation of visuals such as bar plots (clustered and standard), correlation matrices, times series, and scatterplots.

\subsection{Qualitative Analysis}
The qualitative analysis will be conducted on the daily ‘Notes’ provided by the participant. The logs will be analyzed using a content analysis approach to identify keywords, people, or ideas that affect biometric output. The analysis will focus on the words and phrases used by the participant to describe his daily experiences, emotions, and activities. The most common words will be identified and compared with the biometrics from their respective observations. Additionally a sentiment score will be calculated for every word and ‘Notes’ observation. This will be compared to the other biometrics as well. The purpose of this analysis is to gain a deeper understanding of the participant's experiences and identify linguistic trends in relation to HRV and Readiness scores.

\subsection{Programming Languages and Frameworks}
Python will be used for all analysis via the Data Spell IDE. The following packages will be used by general function:

Data Analysis
\begin{itemize}
\item pandas
\item numpy
\item scipy
\item collections
\end{itemize}

Visualization
\begin{itemize}
\item matplotlib
\item seaborn
\item tabulate
\end{itemize}

Text Processing
\begin{itemize}
\item re
\item textblob
\item nltk
\end{itemize}

\section{Results} \label{sec:results}
\subsection{Exploration} \label{subsec:explo}
\begin{figure}[H]
\begin{minipage}[b]{0.49\linewidth}
    \centering
    \includegraphics[width=\linewidth, height=2.6in]{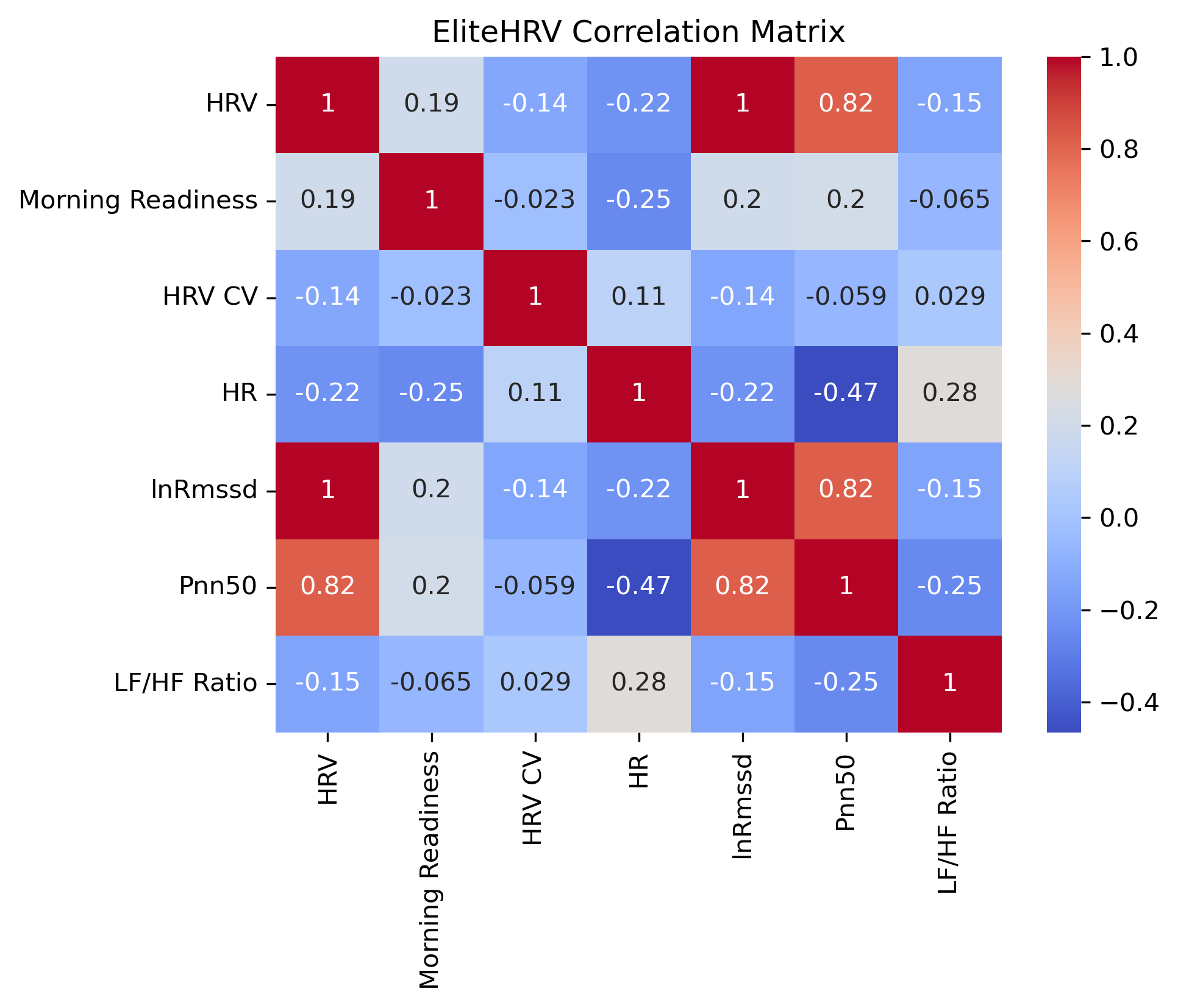}
    \caption{Correlation plot of all relevant heart rate-related variables, including sleep, exercise, and soreness log-entry data.} 
    \label{fig:1}
\end{minipage}
\hfill
\begin{minipage}[b]{0.49\linewidth}
    \centering
    \includegraphics[width=\linewidth, height=2.4in]{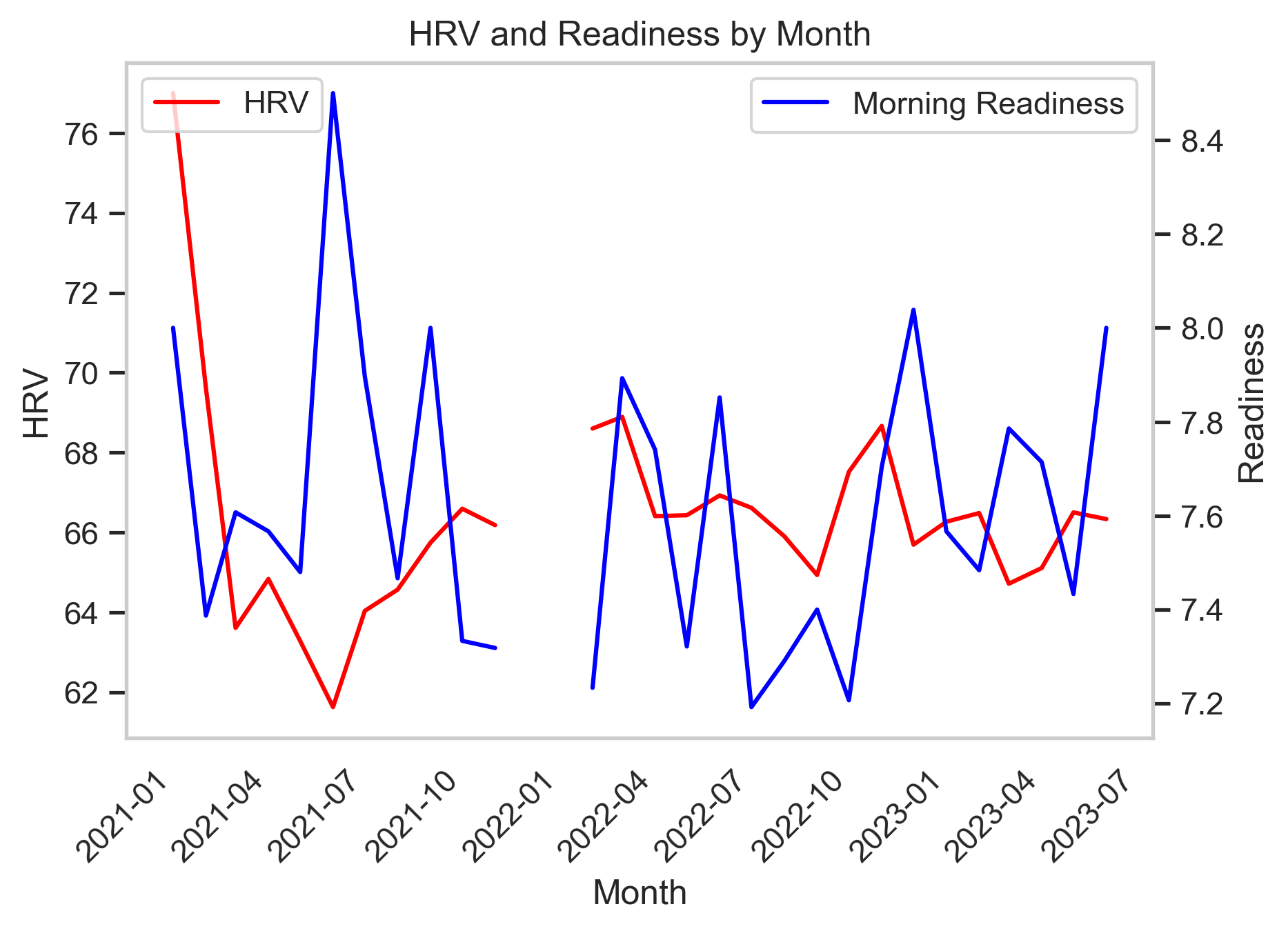}
    \caption{Times series of HRV and Readiness, averaged by month, across the 29 months of data logging. Gap in data appears due to absence of data collection in November and December of 2021 (with the exception of 1 log in November and 3 in late December).}
    \label{fig:2}
\end{minipage}
\end{figure}

\noindent Before beginning any substantial analysis, it is first important to become acquainted with the data. Figure \ref{fig:1} begins this exploration showing seven of the fourteen EliteHRV biometric variables and all corresponding Pearson\footnote{Pearson is used as it assumes normality and a linear relationship between correlates.} correlation coefficients. Only seven were used as many of the variables are only slightly different (or are providing the same information in a different way). For example, LF/HF Ratio captures the same information as both Low Frequency Power and High Frequency Power, by providing the ratio between the two (a typical indicator of ANS balance).

\noindent A few Pearson's coefficients stand out. HRV and Pnn50 have a very strong correlation (r=.82). HRV has a perfect lnRmssd (r = 1.0) indicating these are measuring essentially the same thing. The remainder of the variables have generally moderate correlations, particularly with HRV, indicating it is safe to focus the analytical efforts on this one variable. Given there is a moderate to strong relationship with all other relevant biometrics, identifying what lifestyle choices influence HRV will in turn provide information about how those lifestyle choices will likely impact the rest of the biometrics.

\noindent Figure \ref{fig:2} takes a step back to visualize the scope and touch briefly on the limitations of the data. The time-series plot shows HRV (left axis) and Morning Readiness score (right axis) over time. As is described in the data section, readings from the participant began in late December 2020 up until early May 2023. There is however a clear break in both lines from November to December of 2021 - these months had a combined total of 6 readings so they were omitted from analysis. The plot also illustrates the weak-to-moderate correlation between HRV and Morning Readiness (r = .19) shown in Figure \ref{fig:1}. The lines (for the most part) seem to follow a similar trend with significant exceptions in July 2021, July 2022, and October 2022. Also of note, Morning Readiness appears more sporadic than HRV; however each are on very different scale so this is understandable.

\subsection{Anxiety: Relationship to HRV and Appraisal}
Various studies have found a strong connection between negative fluctuations in HRV and increased anxiety (Kim, 2018). Similar studies have addressed different types of appraisal strategies\footnote{This is particularly true in sports and other competitive settings (Lagos et al., 2008; Jarvelin-Pasanen et al., 2018)}, most commonly, reframing anxiety as excitement. One Harvard Business School study found that "state anxiety lowers self-efficacy" (ie. self-belief) which "in turn, profoundly influences decision making and behavior." This same study found that reappraising anxiety as excitement resulted in statistically significant improvements in said self-efficacy (Brooks, 2014). Knowing this as well as the well-documented, unpleasant sensations that arise as a product of anxiety, it may be helpful to identify the biometrics most highly associated with anxiety so as to target these variables in analysis.

\noindent As was just stated, anxiety serves to alert the body of danger so as to promote survival. The autonomic nervous system is heavily involved in this process, particularly with regards to the sympathetic nervous system (S or SNS). The SNS is responsible for the "fight or flight" responses while the parasympathetic nervous system (PS or PNS) is responsible for "rest and recover" responses. As one might imagine, activation of the sympathetic nervous system and anxiety are both closely related stress-responses (Goldstein, 1987). Knowing this it is essential to identify what biometrics are most different across the PS and S groups.

Looking to Table \ref{tab:1}, with emphasis on the bold values, there are clear differences between each biometric average across the PS and S observations. Most notably, average HRV among S observations is more than 6.5 units lower than that of PS observations. Additionally, average Total Power of the PS group is ~136.7\% larger than that of the S group. Based on these differences, and statistically significant p-values \footnote{Using a two sample t-test, all five biometrics shown in Table \ref{tab:1} yielded a p-value below .001 indicating \textit{strong} statistical significance, with the exception of 'LF/HF Ratio' (p-value = .0075) (still statistically significant). These p-values indicate an ability to reject the null hypothesis and assume the means are different between groups}, it is safe to assume both HRV and Total Power are significantly different between the ANS groups.
\begin{table}[H]
    \centering
    \caption{Mean HRV, LF/HF Ratio, Low Frequency Power, High Frequency Power, and Total Power by autonomic nervous system balance}
    \label{tab:1}
    \begin{tabular}{lcccccc}
    \toprule
        ANS Balance & HRV & LF/HF Ratio & Low Frequency Power & High Frequency Power & Total Power & n \\
        \midrule
        PS & \textbf{69.5241} & 1.73542 & 4543.27 & 2936.29 & \textbf{7479.56} & 353 \\
        S & \textbf{63.0665} & 1.98358 & 2061.08 & 1098.55 & \textbf{3159.63} & 391 \\
        \bottomrule
    \end{tabular}
\end{table}
\noindent Given that the sympathetic nervous system and anxiety are closely intertwined, and that the S observations appeared to have significantly lower HRV and Total Power (and significant differences among the other biometrics), the analysis will shift towards observing the relationship of these variables with self-documented anxiety. In theory, if sympathetic activity is closely associated with anxiety and has a statistically significant relationship with HRV and Total Power, then anxiety should have those same significant relationships. 

\noindent Table \ref{tab:2} contains four types of anxiety appraisals (recall the discussion in the first paragraph of this section) and the average HRV, Readiness, LF/HF Ratio, and Total Power. LF/HF Ratio was included in this table, despite its less significant p-value as there is significant controversy over its connection to ANS balance and thus anxiety (Billman, 2013). Its inclusion will hopefully serve as a contribution to this debate, albeit a small one.

At first glance, it appears the connection between anxiety, the sympathetic nervous system, and the designated biometrics is supported. Looking toward the bold values there is a 1.66 unit difference between HRV when not reappraising as excitement and when reappraising. This seems to show an increase in HRV (which seems to be associated with more parasympathetic activation and less sympathetic activation based on findings from Table \ref{tab:1}). There appears to also be a significant difference between these two groups' Total Power; however, when testing for significance between these two groups using a two-sample t-test, there are no statistically significant differences between the group means except for LF/HF Ratio (p-value = .049). As some speculate, higher values of LF/HF Ratio may be associated with more sympathetic activation. While no statistically significant findings supporting this were found in Table \ref{tab:1}, Table \ref{tab:2} shows a statistically significant increase in LF/HF Ratio in anxious observations when not reappraising as excitement. 
\begin{table}[H]
    \centering
    \caption{Mean HRV, Readiness, and LF/HF Ratio by anxiety appraisal}
    \label{tab:2}
    \begin{tabular}{lccccc}
    \toprule
        Appraisal & HRV & Readiness & LF/HF Ratio & Total Power & n \\
        \midrule
        Anxious & 66.29 & 7.55 & 1.8 & 5396.07 & 287 \\
        Excited & 66.43 & 7.51 & 1.8 & 5335.78 & 389 \\
        \midrule
        Anxious, not Excited & \textbf{64.71} & 7.36 & \textbf{2.42} & \textbf{4871.24} & 14 \\
        not Anxious, Excited & 66.57 & 7.4 & 1.86 & 5130.54 & 116 \\
        Anxious, Excited & \textbf{66.37} & 7.56 & \textbf{1.77} & \textbf{5422.99} & 273 \\
        Neither & 65.45 & 7.7 & 1.94 & 4633.8 & 291 \\
        \bottomrule
    \end{tabular}
\end{table}
\noindent These findings from Table \ref{tab:1} and \ref{tab:2} are important for two reasons:
\begin{enumerate}
    \item There is a statistically significant difference in average LF/HF Ratio between PS and S observations as well as between observations indicated as 'Anxious' but not 'Excited' and those indicated as 'Anxious' and 'Excited'. This may be relevant or useful in the debate outlined in Billman's article (2013).
    \item There seems to be some evidence of effective anxiety reappraisal with regards to change in relevant biometrics. LF/HF Ratio is significantly higher in 'Anxious' observations that indicate no excitement, than those that do indicate excitement. A higher LF/HF ratio is related to sympathetic activation, so this suggests that re-appraisal may reduce sympathetic activation and, in-turn, anxiety \footnote{Please see the \hyperref[subsec:limit]{Limitations} section for details on self-reporting appraisal.}. 
\end{enumerate}
\subsection{Tags} \label{subsec:tags}
While there is little evidence that the participant's self-reported anxiety data is associated with any of the biometrics (except LF/HF Ratio), there is significant evidence that autonomic nervous system state (PS or S) is related to these metrics. Further, connection between anxiety and sympathetic nervous system is well documented, so for the purpose of this study, that association will be assumed to be significant (Hoehn-Saric, 1988; Richards, 2000; Bellocchio et al., 2013; El-Sheikh et al., 2009). Assuming a strong association between sympathetic activation (S) and anxiety, and knowing there is a statistically significant difference in HRV, power, and LF/HF ratio (among other things) between S and PS, it would be useful to delve into the relationships between all biometrics with the rest of the self-recorded variables. However, given HRV's well-studied associations with other health determinants such as cardiovascular fitness and longevity, the rest of the study will focus primarily on HRV (Kim et al., 2018; Reginato, 2020; Shaffer, 2014).  

Figure \ref{fig:3} illustrates the most influential binary Tags (self-designed and self-recorded) based on absolute value of Point Biserial correlation. Point Biserial is used in lieu of the more traditional Pearson's because it does not assume a linear relationship or a normal distribution. This is important because HRV is continuous and each of the Tags is a binary. On top of each bar is the specific correlation coefficient for that Tag as well as the number of appearances in the data (n) and an asterisk(s) denoting significance (*** $<$ .001; ** $<$ .01; * $<$ .05). Statistical significance in this case means that the correlation coefficient was not just due to chance or luck, it does not denote a significant relationship between the two correlates.
\subsubsection{Alcohol and Tobacco}
\noindent The strongest correlates with HRV are unsurprisingly Alcohol (7+ drinks), Alcohol (4-6 drinks), and Veisalgia. While the correlations are only weak to moderate, these negative relationships align with existing research. One study out of the Yale School of Medicine found that individuals with Alcohol Use Disorder suffered from lower HRV than non AUD individuals. They further found that HRV only improved in AUD individuals after 4 months of abstinence (Ralevski et al., 2019). Further research interestingly found that social anxiety in individuals was positively related to "alcohol-related problems", meaning that not only does alcohol lead to anxiety through lowering HRV, but those with anxiety are more likely to have problems from alcohol consumption - a cycle of sorts (Schry \& White, 2013; Norberg et al., 2011).

\noindent Tobacco (smoking) also had a significant relationship with HRV, with a Point Biserial coeffient ($r_{pb}$) of -0.16. However, this relationship should not be considered given there is only a sample size of 1 and there is likely confounding from the alcohol variables.
\subsubsection{Sleep}
\begin{figure}[H]
\begin{minipage}[b]{0.48\linewidth}
    \centering
    \includegraphics[width=\linewidth, height=2.5in]{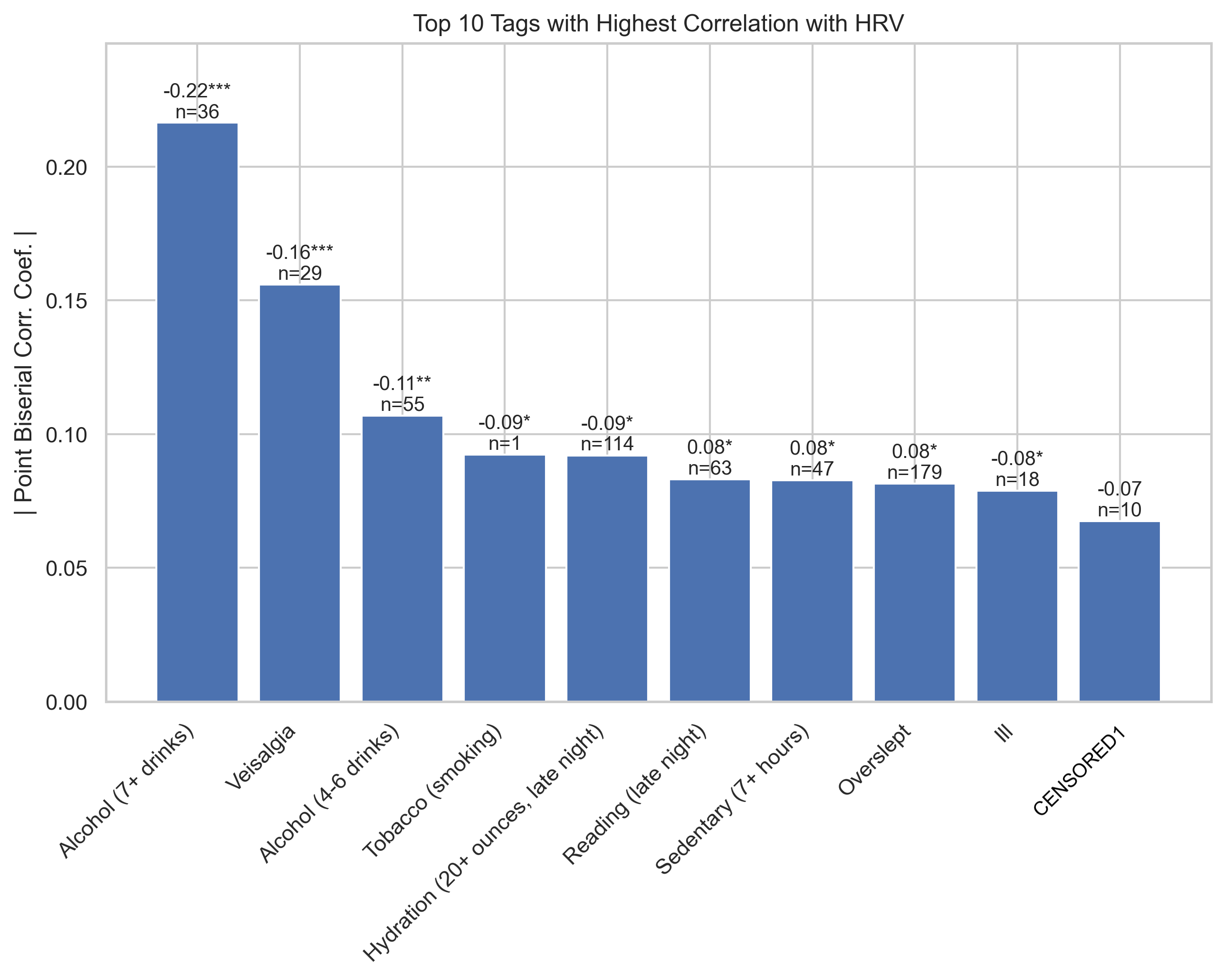}
    \caption{Barplot showing ten binary Tags with the highest Point Biserial correlation (abs. value) with HRV. Asterisks denote statistical significance ($***<.001; **<.01; *<.05$)} 
    \label{fig:3}
\end{minipage}
\hfill
\begin{minipage}[b]{0.48\linewidth}
    \centering
    \includegraphics[width=\linewidth, height=2.5in]{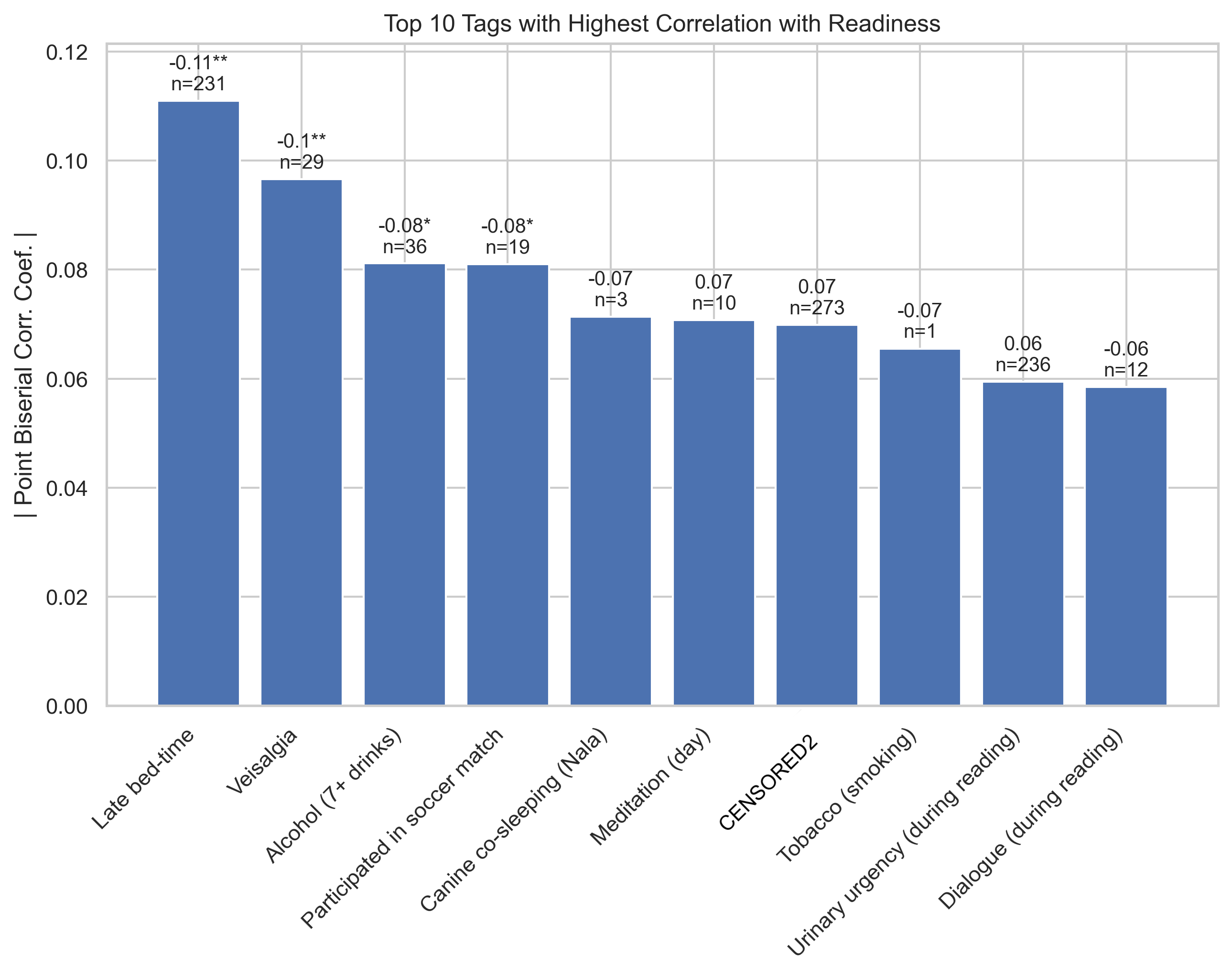}
    \caption{Barplot showing ten binary Tags with the highest Point Biserial correlation (abs. value) with Morning Readiness score. Asterisks denote statistical significance ($***<.001; **<.01; *<.05$)}
    \label{fig:4}
\end{minipage}
\end{figure}

Figure \ref{fig:3} also had a number of sleep-related variables among the most significant correlates. Hydration (20+ ounces, late night), Reading (late night), and Overslept all are related to sleep quality and duration. All had weak but statistically significant coefficients. Consuming over 20 ounces of water within one hour of sleeping had a negative correlation of -.08, as did reading within one hour before sleeping. Additionally, when looking at Figure \ref{fig:4} which shows identical relationships but with Morning Readiness, Canine co-sleeping (Nala) shows up as a weak negative correlation. This indicates that when the participant sleeps in the same bed as their dog, Nala, they have a slightly lower Readiness score; however, this coefficient is not significant, so the correlation could be random chance.

\noindent There is limited research on hydration before bed, however there are many papers describing the detrimental ANS effects of sleep deprivation from bright light exposure before bed or other disturbances (Wallace-Guy et al., 2002; Tobaldini et al., 2017).
\subsubsection{Sedentary}
Another statistically significant correlate, is Sedentary (7+ hours). Sitting for more than 7 hours in a day had a weak negative correlation($r_{pb} = -.08$) with HRV. This relationship is statistically significant so the coefficient is assumed to be non-random. 

\noindent Supporting this, Rebar et al. found in two separate studies that sitting time, specifically while at a computer were associated with higher levels of depression, anxiety, and stress symptoms (2014).
\subsubsection{Other}
Other variables that were among the strongest correlates with HRV and Readiness were Ill, Participated in soccer match, Meditation (Day) \footnote{Regardless of the lack of statistical significance, there is abundant new research on the impact of meditation and mindfulness on HRV that may be of interest (Tang et al., 2009; Nijjar et al., 2014)}, Urinary Urgency (during reading), and Dialogue (during reading). Of these Ill is the only coefficient that yields statistical significance. Understandably it is negatively correlated with HRV. Given the lack of significance for the rest of the variables' $r_{pb}$, they will not be described.

\subsection{Sentiment} \label{subsec:sent}
\begin{figure}[H]
\begin{minipage}[b]{0.48\linewidth}
    \centering
    \includegraphics[width=\linewidth, height=2.5in]{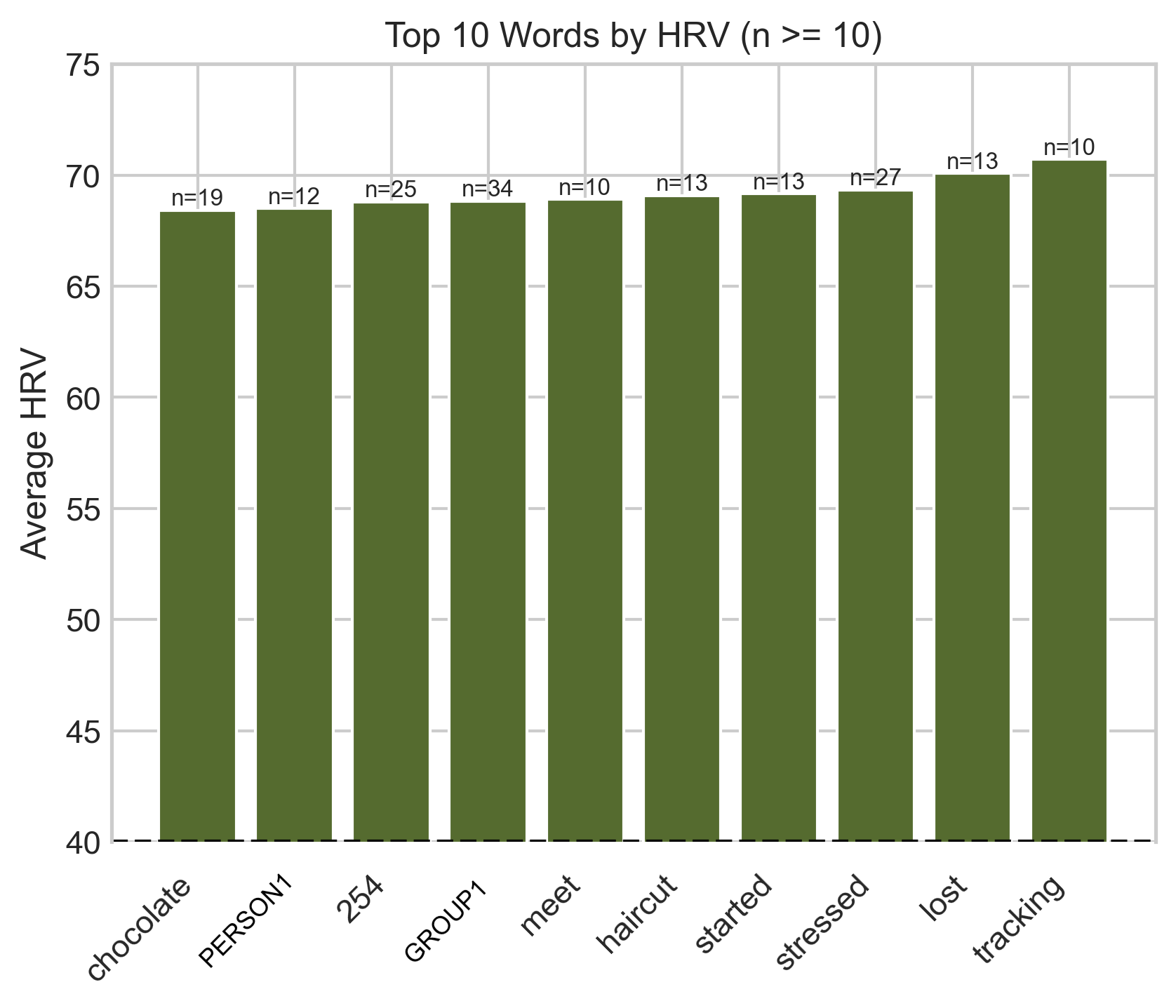}
    \caption{Barplot showing the ten words (with $n>10$) that have the highest average HRV. Words are drawn from each daily log in the 'Notes' variable and are aggregated along with their associated HRV.} 
    \label{fig:5}
\end{minipage}
\hfill
\begin{minipage}[b]{0.48\linewidth}
    \centering
    \includegraphics[width=\linewidth, height=2.5in]{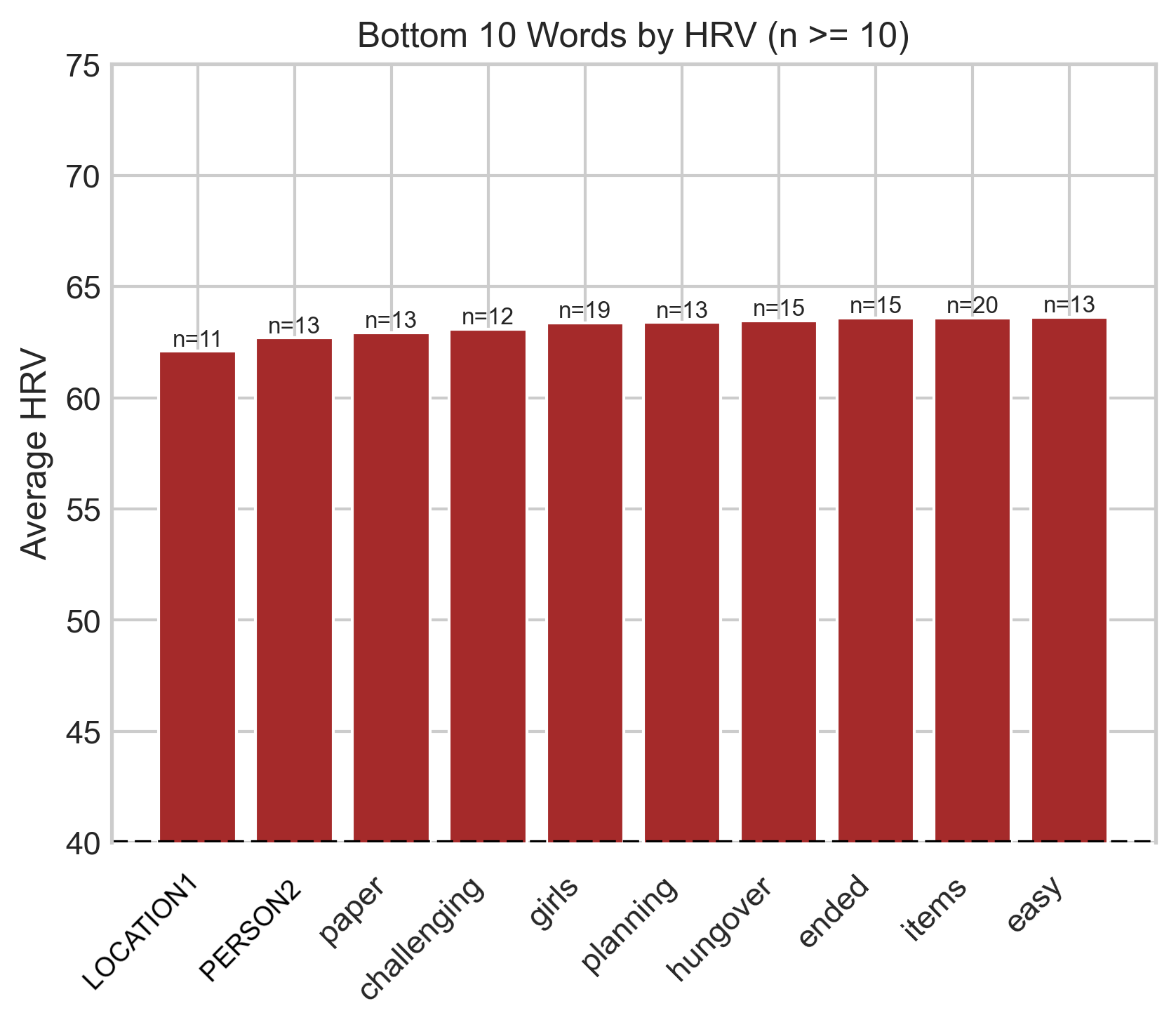}
    \caption{Barplot showing the ten words (with $n>10$) that have the lowest average HRV. Words are drawn from each daily log in the 'Notes' variable and are aggregated along with their associated HRV.}
    \label{fig:6}
\end{minipage}
\end{figure}

\noindent To conclude the analysis, the 'Notes' variable will be accessed and a sentiment score for each daily log will be calculated. In addition the most common words will be tracked and their respective HRV will be aggregated and averaged as is shown in Figures \ref{fig:5} and \ref{fig:6}. This section of the analysis is particularly explorative as results here are difficult to interpret without making assumptions or extrapolating.

\noindent The former of the two figures, Figure \ref{fig:5}, displays the top ten words (occurring more than ten times in the 747 readings) based on average HRV. Some words have expected associations with higher HRV such as chocolate, PERSON1, GROUP1, meet, and started \footnote{For participant-specific context, chocolate is a favorite snack, PERSON1 is a close friend, and GROUP1 is a student-run organization}. Meet and started have subjectively optimistic tones which may factor into the higher HRV. Of note, PERSON1, GROUP1, and in particular, meet, have to do with social networks and social connection.

\noindent Significant research has been done in recent years observing the effects of social connection on health and happiness. A study by Bosle et al. look at similar biometrics revolving around the ANS (including HRV, Low/High Frequency, etc.) and found their to be significant improvements in HRV among adults with a supportive social connection, such as a marriage partner (2022). Other studies have come to similar conclusions finding that lack of social connection, particularly in older adults, is a strong predictor of depression and anxiety (Newman \& Zainal, 2020; Levula et al., 2018).

Some of the other words such as stressed, lost, or tracking have less explainable connections.
\noindent The right-most figure, Figure \ref{fig:6}, conversely shows the ten words (with $n>10$) that have the lowest average HRV. Among them is paper, challenging, girls, hungover, planning, easy, and ended. Interestingly, 'started' was one of the words with the highest HRV and 'ended' has one of the lowest. Also surprising is the presence of PERSON2 and LOCATION1 \footnote{PERSON2 is the name of a close friend of the participant and LOCATION1 is the name of a park where the participant frequently plays soccer} due to the positive associations described by the participant.

\section{Discussion} \label{sec:discussion}
The exploration of the data revealed that HRV had a moderate to strong correlation with all other relevant biometrics, indicating that it is safe to focus the analytical efforts on this one variable. The time-series plot of HRV and Morning Readiness score showed a weak-to-moderate correlation between the two variables. This prompted the inclusion of Readiness in the later stages of analysis. 

\noindent The relationship between anxiety and HRV was explored, with an emphasis on identifying the biometrics that are most highly associated with anxiety. The results showed that HRV and Total Power were significantly different between the autonomic nervous system (ANS) groups. 

\noindent The Tags most closely related to HRV were alcohol, tobacco, sleep, and sedentary behavior. Tobacco (smoking) also had a negative relationship with HRV, although this relationship should not be considered given the small sample size. Hydration (20+ ounces, late night), Reading (late night), and Overslept all had weak but statistically significant coefficients related to sleep quality and duration. Sedentary behavior for more than 7 hours in a day was found to have a weak negative correlation with HRV.

\noindent The sentiment analysis of the participant's notes revealed that words related to social connection were associated with higher HRV. However some words such as lost and tracking had less explainable connections.

\subsection{Limitations}\label{subsec:limit}
\noindent The study has several limitations due to the small sample size, including the lack of statistical significance for some of the results. However, the results provide insight into the potential impact of lifestyle choices on HRV and the autonomic nervous system. The implications of the results suggest that changes in lifestyle, such as reducing alcohol consumption, increasing social connection, and improving sleep habits/hygiene, may lead to improved HRV and in turn reduced anxiety and sympathetic activation. Furthermore, their a few participant-specific limitations:
\begin{itemize}
\item The participant occasionally (consciously or unconsciously) refrained from selecting 'Anxious' in the self-report questionnaire as there was concern that acknowledging this emotion could negatively impact their performance or confidence during an important game or event. Similarly, they may have selected 'Excited' even if they were not truly experiencing that emotion as they believed it would enhance their performance. These actions may introduce bias into the self-report data. This inconsistent data is likely the reasoning behind the lack of statistical significance in Table \ref{tab:2}
\item The participant acknowledges not caring for their heart rate (HR) monitor as meticulously as is recommended by the manufacturer's instructions, which state to rinse and air dry the device after each use. The participant admits to neglecting this maintenance routine since the early months of recording. It is possible that this has led to less reliable readings from the HR monitor (D. Witt et al., n.d.; Coutts et al., 2020; B. Carper et al., 2020).
\item The language and structure used by the participant in recording their daily notes have undergone changes. Specifically, the participant has shifted from a general overview of their day to categorizing their observations as 'Good,' 'Bad,' or 'Target,' as is first observed in entry 258.  
\end{itemize}

\subsection{Future Work}
Interesting future analysis would include more data and would expand from a single-case study to a population of participants. Ideally the time frame would span longer than 26 months with stricter and more consistent data recording practices. Interesting variables or outside research to include in the next iteration are as follows:
\begin{enumerate}
    \item Live tracking of HRV using resonance frequency breathing to access flow-state and mitigate competitive anxiety. This is approaching anxiety and HRV from a different perspective, but existing studies have been able to reduce competitive anxiety and access flow state through monitoring HRV and applying resonance breathing (Steffen et al., 2017).
    \item Track daily nutrition in take beyond Tags about excessive snacking or extreme malnutrition. Logs of unplanned eating (Dikariyanto V et al., 2020) and water and sugar intake would be fascinating additions.
    \item Measurement of social connection or time with others.
    \item Inclusion of mandates positive and negative self-talk before each day or significant event. This data, paired with post-day/event sentiment, could be a fascinating analysis (Hatzigeorgiadis, 2008).
    \item With a study using a population n of participants, include a personality exam prior, during, and after trials. Use these traits to categorize anxiety and patterns in biometrics (LeBlanc, 2017).
\end{enumerate}

\subsection{Ethical Considerations}
The participant has provided consent to use his data for research purposes. The participant's identity will remain anonymous. The author asks that readers refrain from sharing data or independent analysis.

\section{Appendix} \label{sec:appendix}
\subsection{Variables} \label{subsec:variables}
\subsubsection{Biometric} \label{subsubsec:biometric}
\begin{itemize}
    \item \textit{Date of the reading}
    \item \textit{Duration of the reading}
    \item \textit{ANS Balance} which reads S (sympathetic) or PS (parasympathetic) indicating the balance of the autonomic nervous system at the time of reading.
    \item \textit{HRV} Score which reads 1 to 100 and provides a heart rate variability measure (measure of variation in between heart beats, with a higher score indicating better cardiovascular health) at the time of reading.
    \item \textit{lnRMSSD} is the natural logarithm of the root mean squared successive differences between RR signals (RR signals are the measure of R waves in an ECG signal). This is another HRV method, but is less affected by postural and respiratory changes. The natural log makes the distribution of data more normal and thus better for analysis.
    \item \textit{RMSSD} is the root mean squared successive differences between RR signals (RR signals are the measure of R waves in an ECG signal). This is another HRV method, but is less affected by postural and respiratory changes.
    \item \textit{NN50} is the number of pairs of adjacent RR intervals that differ by more than 50 milliseconds. This is a simpler measure of HRV. 
    \item \textit{PNN50} is the percentage of pairs of adjacent RR intervals that differ by more than 50 milliseconds. This is a simpler measure of HRV.
    \item \textit{SDNN} is the standard deviation of all NN intervals (time in between successive R waves in an ECG reading). This is a simpler measure of HRV. 
    \item \textit{7 Day Average of HRV}
    \item \textit{7 Day Average of HRV CV} CV is the average over seven days of the coefficient of variation which is the ratio of the standard deviation and mean. It describes the variability of the HRV measures themselves, with a higher score indicating more variance.
    \item \textit{HR} is heart rate in beats per minute.
    \item \textit{7 Day Average HR}
    \item \textit{Total Frequency Power} is the power in Hz released by both the low frequency (associated with sympathetic activity) and high frequency (associated with parasympathetic activity) bands.
    \item \textit{Low Frequency Power} is the power in Hz released by the low frequency (associated with sympathetic activity).
    \item \textit{High Frequency Power} is the power in Hz released by the high frequency (associated with parasympathetic activity).
    \item \textit{LF/HF Ratio} is the ratio of low frequency power to high frequency power. A score greater than 1 indicates an imbalance towards the sympathetic nervous system, less than 1 indicates an imbalance towards the parasympathetic nervous system, and equal to 1 indicates perfect autonomic balance. For better descriptions of the frequency variables, please see the Elite HRV description.
    \item \textit{Morning Readiness} is a measure calculated by EliteHRV. This is a number 1-10 that assesses ‘readiness’ based on autonomic changes. A low score indicates a big change from parasympathetic to sympathetic (or vice versa). A high score indicates little change in HRV. This score is calculated to improve “long term” performance, not day-of performance (according to Elite HRV)
    
\end{itemize}
\subsubsection{Self-measured} \label{subsubsec:self}

\begin{itemize}
    \item \textit{CENSORED1 (day)} 
    \item \textit{CENSORED1 (night)}
    \item \textit{Alcohol (1-3 drinks)}
    \item \textit{Alcohol (4-6 drinks)}
    \item \textit{Alcohol (7+ drinks)}
    \item \textit{Anxious} describes feeling anxious, nervous or stressed.
    \item \textit{Aroused} indicates being sexually aroused.
    \item \textit{Bruised toe}
    \item \textit{Constipated}
    \item \textit{Different sleeping arrangement} is for when sleeping in a new location. Only new for three days, then the next switch in arrangement is new.
    \item \textit{Excited} indicates feeling eager and excited.
    \item \textit{Soreness (gluteus medius)}
    \item \textit{Headache} 
    \item \textit{Veisalgia} consists of a headache, stomach ache, or other discomfort from alcohol consumption.
    \item \textit{Injured} that is preventing participant from doing some range of activity.
    \item \textit{Screen time (late night)} indicates use of phone or computer for more than 5-minutes within an hour of bedtime.
    \item \textit{Hydration (20+ ounces, late night)}
    \item \textit{Soreness (low back)}
    \item \textit{Soreness (low back, gluteus medius)}
    \item \textit{CENSORED2 (before reading)}
    \item \textit{CENSORED2 (day)}
    \item \textit{CENSORED2 (morning)}
    \item \textit{CENSORED2 (night, late night)}
    \item \textit{Participated in soccer match} for at least 30 minutes of an official 90 minute game.
    \item \textit{Meditation (day)}
    \item \textit{Meditation (morning)}
    \item \textit{Meditation (night)}
    \item \textit{Meditating (during reading)}
    \item \textit{Psilocybin}
    \item \textit{Urinary urgency (during reading)} indicates a need to urinate.
    \item \textit{Bowel urgency (during reading)} indicates a need to defecate.
    \item \textit{Different living arrangement} is for when living in a new location (only "new" for seven days, then the next switch in arrangement is new.)
    \item \textit{Nightmare}
    \item \textit{On phone (before reading)} is for when participant uses phone for something other than weather or clock for more than 5-minutes.
    \item \textit{Malnutrition} indicates eating a non-balanced, sugar-dominated meal twice or more in one day.
    \item \textit{Reading (during reading)}
    \item \textit{Reading (late night)}
    \item \textit{CENSORED3 (morning, day)}
    \item \textit{CENSORED3 (night, late night)}
    \item \textit{Ill} indicates having a sickness that causes discomfort doing everyday activities.
    \item \textit{Sedentary (7+ hours)}
    \item \textit{Overslept} is when the participant slept past their alarm.
    \item \textit{Tobacco (smoking)}
    \item \textit{Unplanned eating (day)} is snacking.
    \item \textit{Unplanned eating (night)} is snacking.
    \item \textit{Soreness (hamstring)}
    \item \textit{Gastric pain} is a stomach ache.
    \item \textit{Dialogue (during reading)} is speaking to another person.
    \item \textit{Thirsty}
    \item \textit{Up and about (before reading)} means getting out of bed and move around on two feet.
    \item \textit{Early wake-up} is more than two hours earlier than the seven day average.
    \item \textit{Late bed-time} is more than two hours later than the seven day average.
    \item \textit{Watching Reading} is observing the reading transpire on the phone screen as it happens.
    \item \textit{Television (during reading)} indicates watching TV.
    \item \textit{Cannabis (ingestion, smoking; night, late night)}
    \item \textit{Cannabis (ingestion, smoking; night, late night)}
    \item \textit{Cannabis (ingestion, smoking; morning, day)}
    \item \textit{CENSORED4}
    \item \textit{Canine co-sleeping (PET1)} is sleeping in the same bed as the participants dog, PET1.
    \item \textit{Canine co-sleeping (PET2)} is sleeping in the same bed as the participants dog, PET2.
    \item \textit{Reading Notes} are self recorded notes that are structured openly by the app. I use these notes to record at least one good thing that happened the day prior (preceded by the sun emoji), at least one difficult thing that happened the day prior (preceded by the cloud emoji), and at least one goal for the day (preceded by (the bullseye emoji)). 
    \item Other self recorded data including: \textit{mood}, \textit{illness} (type and severity), \textit{exercise} (type, duration, and intensity).

    \indent \indent \textbf{Key} \\
    \indent \textit{Morning}: wake-up to 11:59am. \\
    \indent \textit{Day}: 12pm to 4:59pm. \\
    \indent \textit{Night}: 5pm to 1-hour before bed time. \\
    \indent \textit{Late Night}: 1-hour before bed-time to bed-time. \\
    \indent \textit{Before Reading}: day of the reading, before the reading. \\ 
    \indent \textit{During Reading}: during the reading. \\
\end{itemize}
\subsection{Supplemental Figures}

\begin{figure}[H]
  \centering
  \includegraphics[width=.5\linewidth]{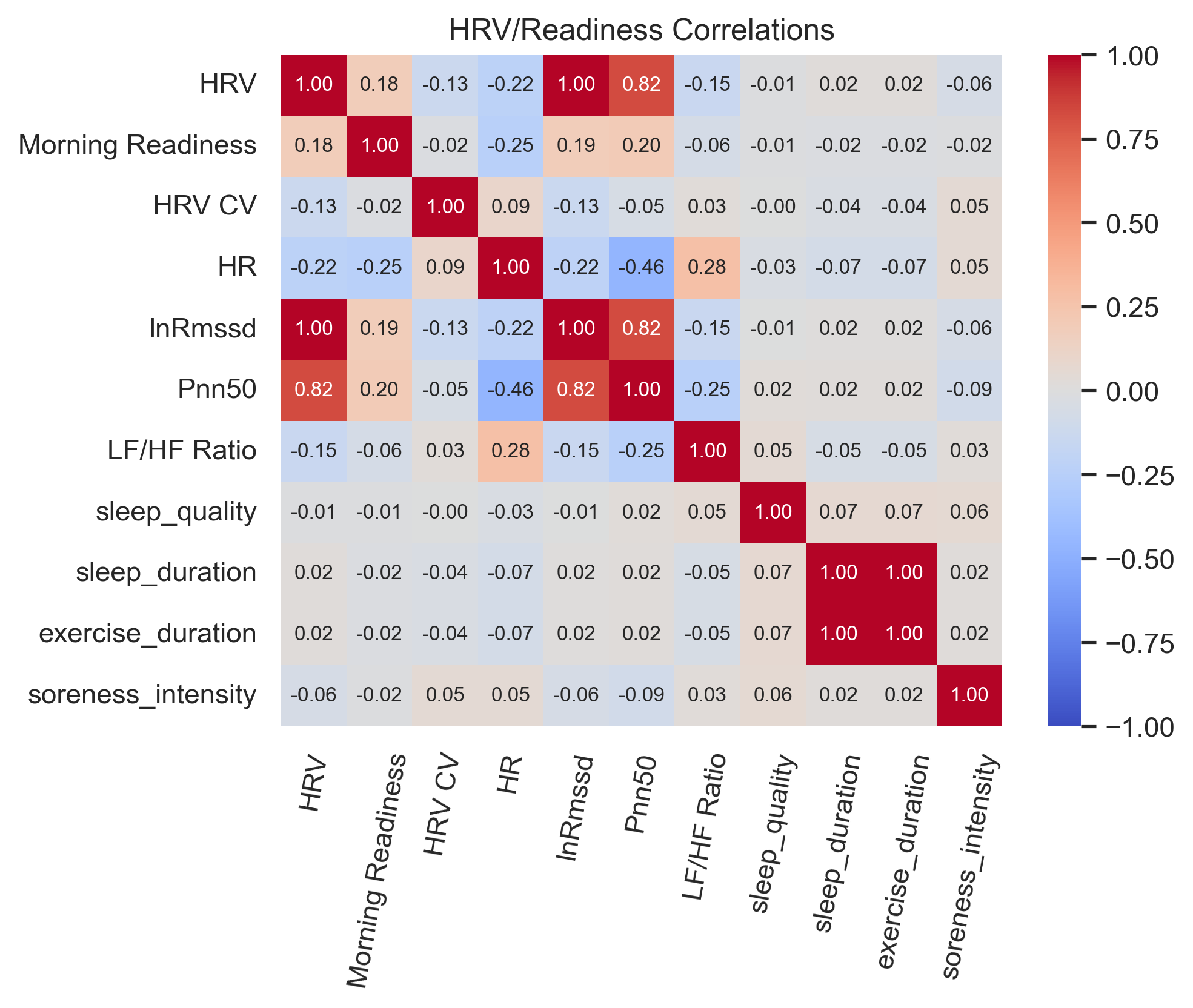}
  \caption{EliteHRV biometrics with addition of sleep, exercise, and soreness variables.}
  \label{fig:updated-correlation}
\end{figure}

\begin{figure}[H]
  \centering
  \includegraphics[width=.5\linewidth]{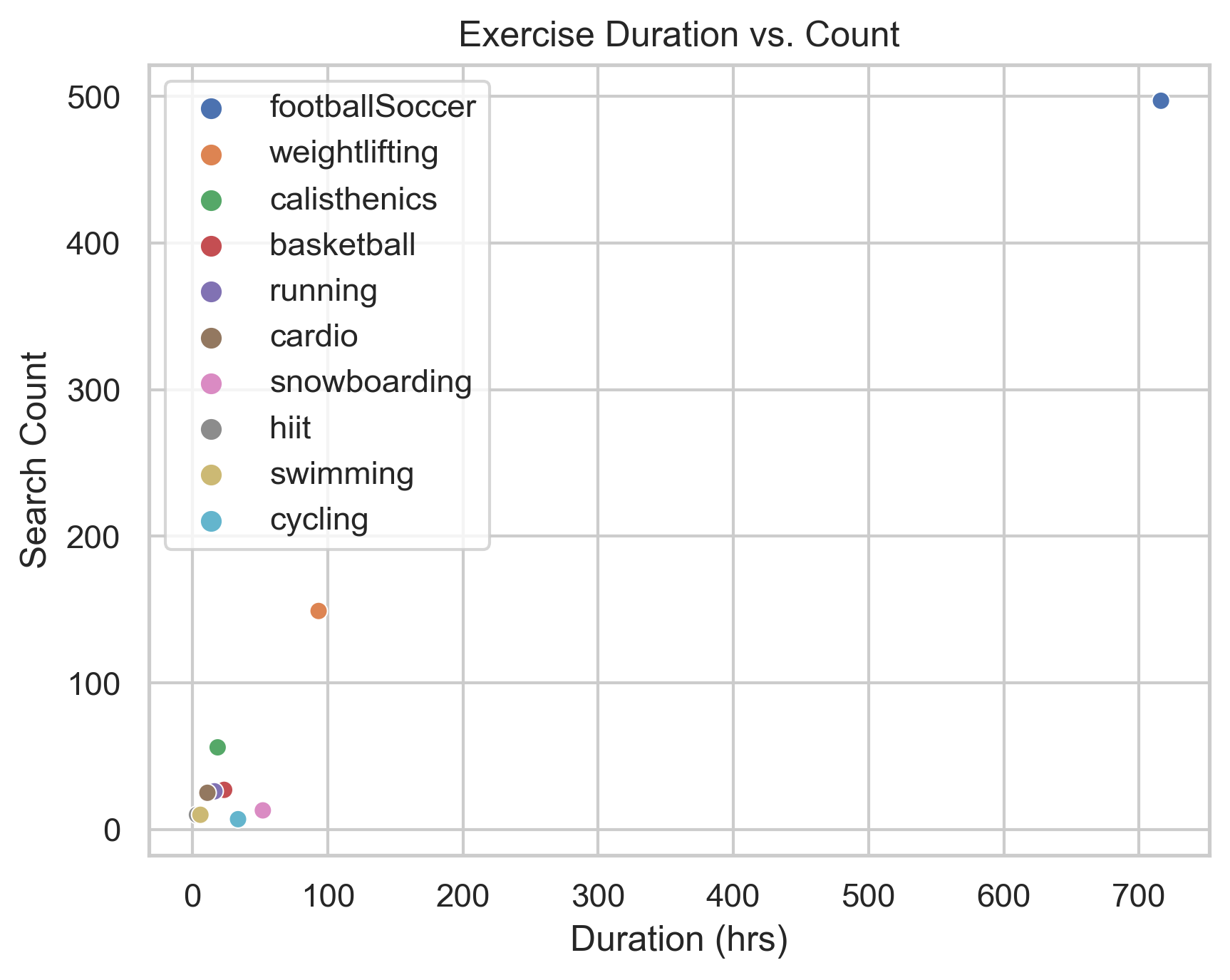}
  \caption{Summary of different types of exercise across 26 months.}
  \label{fig:exercise-summary}
\end{figure}

\section{References}
\begin{enumerate}

    \item B. Carper et al., "Modeling Biological Rhythms to Predict Mental and Physical Readiness," 2020 Systems and Information Engineering Design Symposium (SIEDS), Charlottesville, VA, USA, 2020, pp. 1-6, doi: 10.1109/SIEDS49339.2020.9106683

    \item Bateson, M., Brilot, B., \& Nettle, D. (2011). Anxiety: An evolutionary approach. Canadian Journal of Psychiatry, 56(12), 707-715. doi: 10.1177/070674371105601202
    
    \item Bellocchio, L., Soria-Gómez, E., Quarta, C., Metna-Laurent, M., Cardinal, P., Binder, E., ... \& Marsicano, G. (2013). Activation of the sympathetic nervous system mediates hypophagic and anxiety-like effects of CB1 receptor blockade. Proceedings of the National Academy of Sciences, 110(12), 4786-4791. https://doi.org/10.1073/pnas.1218573110

    \item Billman, G. E. (2013). The LF/HF ratio does not accurately measure cardiac sympatho-vagal balance. Frontiers in Physiology, 4, 26. doi: 10.3389/fphys.2013.00026.
    
    \item Bosle, C., Brenner, H., Fischer, J.E. et al. The association between supportive social ties and autonomic nervous system function—differences between family ties and friendship ties in a cohort of older adults. Eur J Ageing 19, 263–276 (2022). https://doi.org/10.1007/s10433-021-00638-2
    
    \item Brooks A. W. (2014). Get excited: reappraising pre-performance anxiety as excitement. Journal of experimental psychology. General, 143(3), 1144–1158. https://doi.org/10.1037/a0035325

    \item Coutts, L. V., Plans, D., \& Brown, A. W. (2020). Deep learning with wearable based heart rate variability for prediction of mental and general health. Journal of Biomedical Informatics, 112, 103610. https://doi.org/10.1016/j.jbi.2020.103610

    \item D. Witt, R. Kellogg, M. Snyder, J. Dunn, Windows Into Human Health Through Wearables Data Analytics, Current Opinion in Biomedical Engineering, https://doi.org/10.1016/ j.cobme.2019.01.001.
    
    \item Dikariyanto V, Smith L, Chowienczyk PJ, Berry SE, Hall WL. Snacking on Whole Almonds for Six Weeks Increases Heart Rate Variability during Mental Stress in Healthy Adults: A Randomized Controlled Trial. Nutrients. 2020; 12(6):1828. https://doi.org/10.3390/nu12061828

    \item El-Sheikh, M., Kouros, C. D., Erath, S., Cummings, E. M., Keller, P., \& Staton, L. (2009). Marital conflict and children’s externalizing behavior: pathways involving interactions between parasympathetic and sympathetic nervous system activity. Monographs of the Society for Research in Child Development, 74(1), vii–79. doi:10.1111/j.1540-5834.2009.00501.x.

    \item Goldstein D. S. (1987). Stress-induced activation of the sympathetic nervous system. Bailliere's clinical endocrinology and metabolism, 1(2), 253–278. https://doi.org/10.1016/s0950-351x(87)80063-0

    \item Hatzigeorgiadis, A., \& Biddle, S. J. H. (2008). Negative self-talk during sport performance: Relationships with pre-competition anxiety and goal-performance discrepancies. Journal of Sport Behavior, 31(3), 237-257.

    \item Hoehn-Saric, R., \& McLeod, D. R. (1988). The peripheral sympathetic nervous system: its role in normal and pathologic anxiety. Psychiatric Clinics of North America, 11(2), 375-386. https://doi.org/10.1016/S0193-953X(18)30504-5
    
    \item Järvelin-Pasanen, S., Sinikallio, S., \& Tarvainen, M. P. (2018). Heart rate variability and occupational stress-systematic review. Industrial health, 56(6), 500–511. https://doi.org/10.2486/indhealth.2017-0190

    \item Kim, H.-G., Cheon, E.-J., Bai, D.-S., Lee, Y. H., \& Koo, B.-H. (2018). Stress and Heart Rate Variability: A Meta-Analysis and Review of the Literature. Psychiatry Investigation, 15(3), 235-245. https://doi.org/10.30773/pi.2017.08.17
    
    \item Lagos, L., Vaschillo, E., Vaschillo, B., Lehrer, P., Bates, M., \& Pandina, R. (2008). Heart rate variability biofeedback as a strategy for dealing with competitive anxiety: A case study. Biofeedback, 36(3), 109-115.

    \item LeBlanc, J., Ducharme, M. B., \& Thompson, M. (2004). Study on the correlation of the autonomic nervous system responses to a stressor of high discomfort with personality traits. Physiology \& behavior, 82(4), 647-652. doi:10.1016/j.physbeh.2004.05.014
    
    \item Levula, A., Harré, M., \& Wilson, A. (2018). The Association Between Social Network Factors with Depression and Anxiety at Different Life Stages. Community Mental Health Journal, 54(6), 842-854. doi: 10.1007/s10597-017-0195-7

    \item Newman, M. G., \& Zainal, N. H. (2020). The value of maintaining social connections for mental health in older people. The Lancet Public Health, 5(1), e12-e13. doi: 10.1016/S2468-2667(19)30253-1
    
    \item Nijjar, P. S., Puppala, V. K., Dickinson, O., Duval, S., Duprez, D., Kreitzer, M. J., \& Benditt, D. G. (2014). Modulation of the autonomic nervous system assessed through heart rate variability by a mindfulness based stress reduction program. International Journal of Cardiology. doi:10.1016/j.ijcard.2014.08.116
        
    \item Norberg, M. M., Olivier, J., Alperstein, D. M., Zvolensky, M. J., \& Norton, A. R. (2011). Adverse consequences of student drinking: The role of sex, social anxiety, drinking motives. Addictive Behaviors, 36(8), 821-828. https://doi.org/10.1016/j.addbeh.2011.03.010.

    \item Ralevski, E., Petrakis, I., \& Altemus, M. (2019). Heart rate variability in alcohol use: A review. Pharmacology, Biochemistry and Behavior, 176, 83-92.

    \item Rebar, A. L., Duncan, M. J., Short, C., \& Vandelanotte, C. (2014). Differences in health-related quality of life between three clusters of physical activity, sitting time, depression, anxiety, and stress. BMC Public Health, 14(1), 1088. doi: 10.1186/1471-2458-14-1088
    
    \item Rebar, A. L., Vandelanotte, C., van Uffelen, J., Short, C., \& Duncan, M. J. (2014). Associations of overall sitting time and sitting time in different contexts with depression, anxiety, and stress symptoms. Mental Health and Physical Activity, 7(2), 105-110. doi: 10.1016/j.mhpa.2014.02.004

    \item Reginato E, Azzolina D, Folino F, Valentini R, Bendinelli C, Gafare CE, Cainelli E, Vedovelli L, Iliceto S, Gregori D, Lorenzoni G. Dietary and Lifestyle Patterns are Associated with Heart Rate Variability. Journal of Clinical Medicine. 2020; 9(4):1121. https://doi.org/10.3390/jcm9041121

    \item Richards, J. C., \& Bertram, S. (2000). Anxiety sensitivity, state and trait anxiety, and perception of change in sympathetic nervous system arousal. Journal of Anxiety Disorders, 14(4), 413-427. https://doi.org/10.1016/S0887-6185(00)00031-1
    
    \item Schry, A. R., \& White, S. W. (2013). Understanding the relationship between social anxiety and alcohol use in college students: A meta-analysis. Addictive Behaviors, 38(9), 2690-2706.

    \item Shaffer, F., McCraty, R., Zerr, C. L. (2014). A healthy heart is not a metronome: an integra- tive review of the heart’s anatomy and heart rate variability. Frontiers in psychology, 5, 1040. doi: 10.3389/fpsyg.2014.01040
    
    \item Steffen, P. R., Austin, T., DeBarros, A., \& Brown, T. (2017). The Impact of Resonance Frequency Breathing on Measures of Heart Rate Variability, Blood Pressure, and Mood. Frontiers in public health, 5, 222. doi: 10.3389/fpubh.2017.00222
    
    \item Tang, Y.-Y., Ma, Y., Fan, Y., Feng, H., Wang, J., Feng, S., Lu, Q., Hu, B., Lin, Y., Li, J., Zhang, Y., Wang, Y., Zhou, L., \& Fan, M. (2009). Central and autonomic nervous system interaction is altered by short-term meditation. Proceedings of the National Academy of Sciences of the United States of America, 106(22), 8865-8870. doi:10.1073/pnas.0904031106
    
    \item Tobaldini, E., Costantino, G., Solbiati, M., Cogliati, C., Kara, T., Nobili, L., \& Montano, N. (2017). Sleep, sleep deprivation, autonomic nervous system and cardiovascular diseases. Neuroscience \& Biobehavioral Reviews, 74, 321-329. doi:10.1016/j.neubiorev.2016.07.004

    \item Wallace-Guy, G. M., Kripke, D. F., Jean-Louis, G., Langer, R. D., Elliott, J. A., \& Tuunainen, A. (2002). Evening light exposure: implications for sleep and depression. Journal of the American Geriatrics Society, 50(4), 738-739. https://doi.org/10.1046/j.1532-5415.2002.50171.x

\end{enumerate}
\end{document}